\documentclass{webofc}
\usepackage[varg]{txfonts}   
\usepackage{placeins}

\newcommand{\NICA}{{NICA}}
\newcommand{\MPD}{{MPD}}
\newcommand{\hi}{heavy-ion\ }

\newcommand{\TPCfull}{Time-Projection Chamber}
\newcommand{\TPC}{TPC}
\newcommand{\TOFfull}{Time-of-Flight}
\newcommand{\TOF}{TOF}
\newcommand{\FHCal}{{FHCal}}
\newcommand{\UrQMD}{{UrQMD}}
\newcommand{\LAQGSM}{{LAQGSM}}
\newcommand{\GEANT}{{GEANT}}
\newcommand{\PID}{{PID}}
\begin{document}
%
%
\title{Performance studies of anisotropic flow with MPD at NICA}
%
%

\author{\firstname{Petr} \lastname{Parfenov}\inst{1,2}\fnsep\thanks{\email{terrylapard@gmail.com}} \and
        \firstname{Arkadiy} \lastname{Taranenko}\inst{1} \and
        \firstname{Ilya} \lastname{Selyuzhenkov}\inst{1,3} \and
        \firstname{Peter} \lastname{Senger}\inst{1,3}
}

\institute{National Research Nuclear University MEPhI, Moscow, Russia 
\and
           Institute for Nuclear Research of the Russian Academy of Sciences, Moscow, Russia
\and
           GSI Helmholtzzentrum f{\"u}r Schwerionenforschung, Darmstadt, Germany
          }

\abstract{%
  The Multi-Purpose Detector (MPD) at NICA collider has a substantial discovery potential concerning the exploration of the QCD phase diagram in the region of high net-baryon densities and moderate temperatures. The anisotropic transverse flow is one of the key observables to study the properties of dense matter created in heavy-ion collisions. The MPD performance for anisotropic flow measurements is studied with Monte-Carlo simulations of gold ions at NICA energies $\sqrt{s_{NN}}=4-11$ GeV using different heavy-ion event generators. Different combinations of the MPD detector subsystems are used to investigate the possible systematic biases in flow measurements, and to study effects of detector azimuthal non-uniformity. The resulting performance of the MPD for flow measurements is demonstrated for directed and elliptic flow of identified charged hadrons as a function of rapidity and transverse momentum in different centrality classes.
}
\maketitle
\FloatBarrier
\section{Introduction}
\label{intro}
Experimental and theoretical studies of the thermodynamical properties of quark-gluon
matter are one of the top priorities worldwide in high-energy \hi physics \cite{Czopowicz:2012ey}. 
Transverse anisotropic flow measurements are one of the key methods to study the time evolution of the strongly interacting medium formed in nuclear collisions.
In non-central collisions, the initial spatial anisotropy results in an azimuthally anisotropic emission of particles.
The magnitude of the anisotropic flow can be defined via the Fourier coefficients $v_n\{\Psi_{m}\}$ of azimuthal distribution of the emitted particles with respect to the reaction plane \cite{Poskanzer:1998yz}:
\begin{eqnarray}
\frac{dN}{d(\varphi - \Psi_{m})} = \frac{1}{2\pi} \left( 1+2\sum\limits_{n=1}^{\infty}v_{n}\cos \left[ n(\varphi - \Psi_{m}) \right] \right),
\end{eqnarray}
where $\varphi$ -- is the azimuthal angle of the particle, $n$ -- is the harmonic order and $\Psi_{m}$ is the $m$-th order collision symmetry plane angle. $v_1$ and $v_2$ are called directed and elliptic flow, respectively.

\FloatBarrier
\section{Simulation and analysis setup}
\label{sec-setup}

The main workflow is similar to the one described in the previous work \cite{Svintsov:2017rac,Parfenov:2017rnj}.
Ultra-relativistic Quantum Molecular Dynamics (\UrQMD) \cite{Bleicher:1999xi,Bass:1998ca}, and the Los Alamos Quark-Gluon String Model (\LAQGSM) \cite{Gudima:2001aa,Mashnik:2002uj} were used to generate 4M and 100k events, respectively. 
The detector response was simulated using \GEANT\ toolkit (version 3 and 4 - see in sec. \ref{subsec-resolution}).
The resulting signals from the detector subsystems were used as input information for the reconstruction procedure.

The events were divided into classes of centrality ranging from $0$ to $100$\% in steps of 5\%.
In this work, the centrality determination is based on the multiplicity of the charged particles reconstructed by the \TPCfull\ (\TPC), and the anisotropic flow analysis for $Au+Au$ collisions is presented for the two energies corresponding the highest and lowest ones of the \NICA\ collider.

Tracks are selected according to the following criteria: $|\eta|<1.5$; $0.2<p_T<2\ \text{GeV/c}$; $N_{hits}^{TPC}>32$; $ 2 \sigma $ DCA cut for primary particle selection; particle identification (\PID) using $dE/dx$ from \TPC\ and $m^2$ calculated from \TOFfull\ detector (\TOF) \cite{Mudrokh:2017gbk}. 
Where DCA is the distance of the closest approach between the reconstructed vertex and a charged particle track.

\FloatBarrier
\section{Results}
\label{sec-results}

For the collective flow measurement, the event plane method was used \cite{Poskanzer:1998yz}. The reaction plane was estimated from the energy deposition of the nuclear fragments in backward and forward rapidities in the forward hadron calorimeters (\FHCal). $Q$-vector and event plane angle $\Psi_m^{EP}$ were calculated as follows:

\begin{eqnarray}
q_x^{m}=\frac{\sum E_i \cos m\varphi_i}{\sum E_i},\ q_y^{m}=\frac{\sum E_i \sin m\varphi_i}{\sum E_i},\ \Psi_m^{EP}=\textrm{TMath::ATan2}(q_y^m,q_x^m),
\end{eqnarray}
where $E_i$ is the energy deposition in the $i$-th module of \FHCal, and $\varphi_i$ its azimuthal angle. For $m = 1$, the weights had opposite signs for backward and forward rapidities due to the antisymmetry of the $v_1$ as a function of rapidity.
The anisotropic flow $v_n\{\Psi_{m}\}$ is calculated as follows:
\begin{eqnarray}
v_n\{\Psi_{m}\}=\frac{\left\langle\cos\left[n(\varphi - \Psi_m^{EP})\right]\right\rangle}{R_n\{\Psi_m\}},\ R_n\{\Psi_m\}=\left\langle\cos\left[n(\Psi_m^{EP}-\Psi_{m})\right]\right\rangle,
\end{eqnarray}
where $R_n\{\Psi_m\}$ is the event plane resolution, and $\Psi_{m}$ is the $m$-th order collision symmetry plane, which cannot be measured experimentally.
To estimate the event plane resolution, the 2-subevent method with extrapolation
algorithm was used \cite{Poskanzer:1998yz}.
In this work $v_1$ and $v_2$ were measured with respect to the $1$-st order collision symmetry plane $\Psi_1$.

\FloatBarrier
\subsection{Resolution correction factor}
\label{subsec-resolution}

\begin{figure*}
\centering
	\includegraphics[width=1.\textwidth,clip]{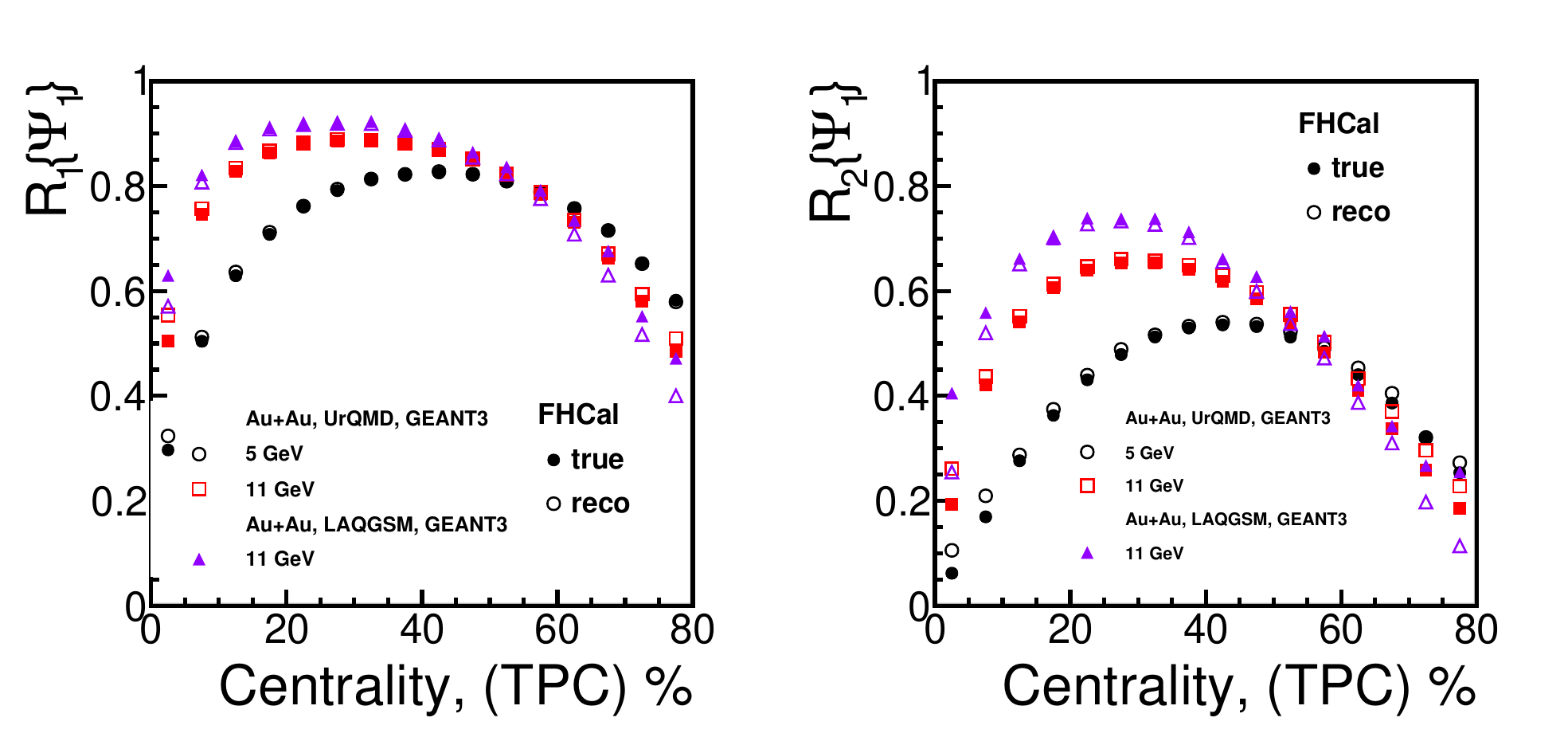}
	\caption{Resolution correction factor as a function of centrality for $v_1$ (left) and $v_2$ (right) for the \UrQMD\ and \LAQGSM\ event generators. The results from the \GEANT 3 simulation marked as true and one from the reconstruction procedure are marked as reco.}
	\label{fig-resolution-all}       
\end{figure*}

In Figure \ref{fig-resolution-all} the resolution correction factor for $v_1$ and $v_2$ is shown. The difference between \LAQGSM\ and \UrQMD\ is due to the fact, that the first model simulates nuclear fragments, while in the second one the spectators consist of neutrons and protons only. Hence, in the case of \LAQGSM, more particles go through the beam hole in the center of \FHCal\ detectors. Apart from that, the results show good performance in a wide centrality range from $0$ to $80\%$ for all energies.

\begin{figure*}
	\centering
	\includegraphics[width=1.\textwidth,clip]{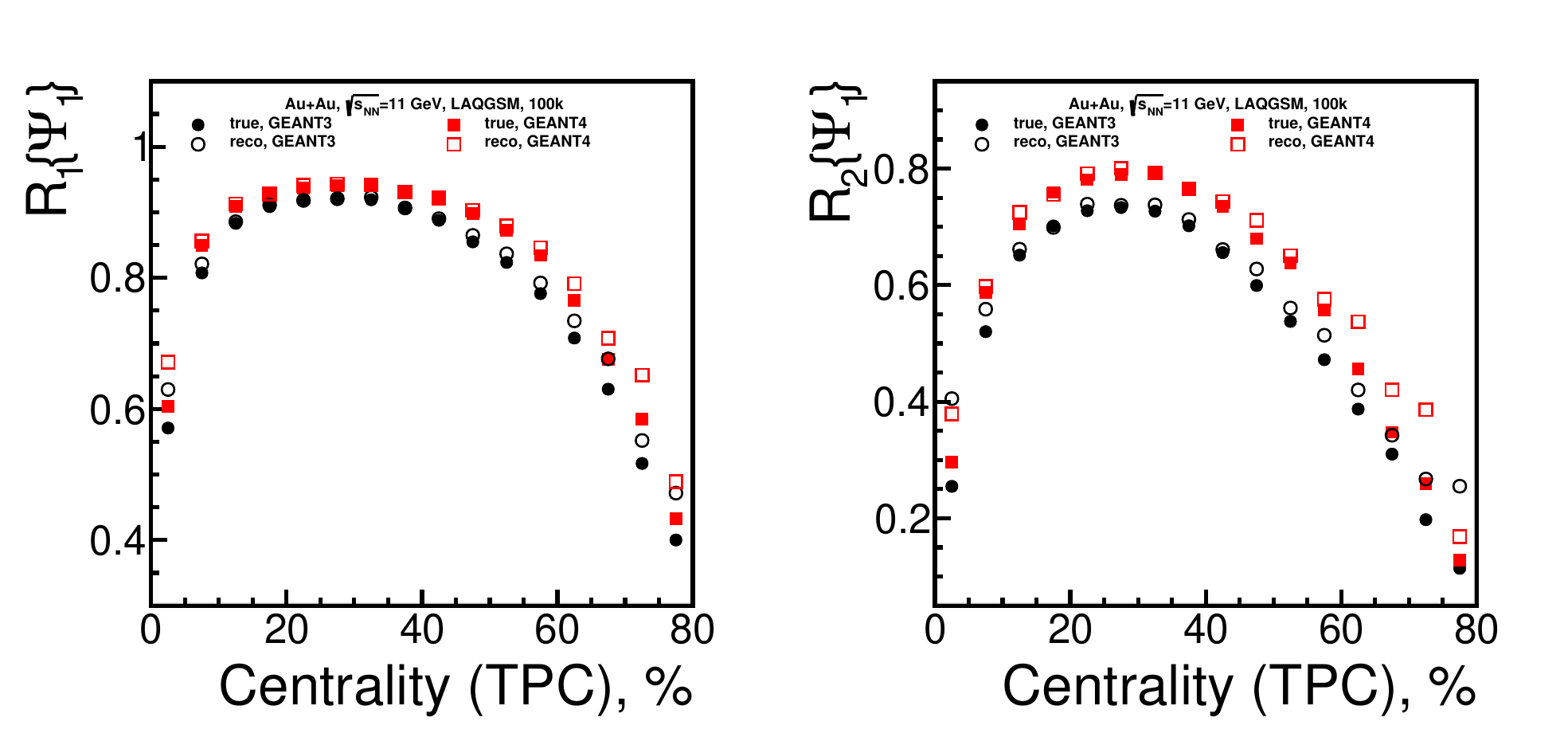}
	\caption{Resolution correction factor as a function of centrality for $v_1$ (left) and $v_2$ (right) for the \LAQGSM\ event generator using \GEANT 3 (\GEANT 4) framework marked as black (red) dots.
	}
	\label{fig-resolution-geant}       
\end{figure*}

In Figure \ref{fig-resolution-geant} the resolution correction factor for both flow harmonics using \LAQGSM\ is presented. At $\sqrt{s_{NN}}=11$ GeV, the energy deposition using \GEANT 4 increases compared to the values given from \GEANT 3 which can be a result of a more realistic hadronic shower simulation.

\FloatBarrier
\subsection{Azimuthal anisotropic flow}
\label{subsec-flow}

\begin{figure*}
	\centering
	\includegraphics[width=0.49\textwidth,clip]{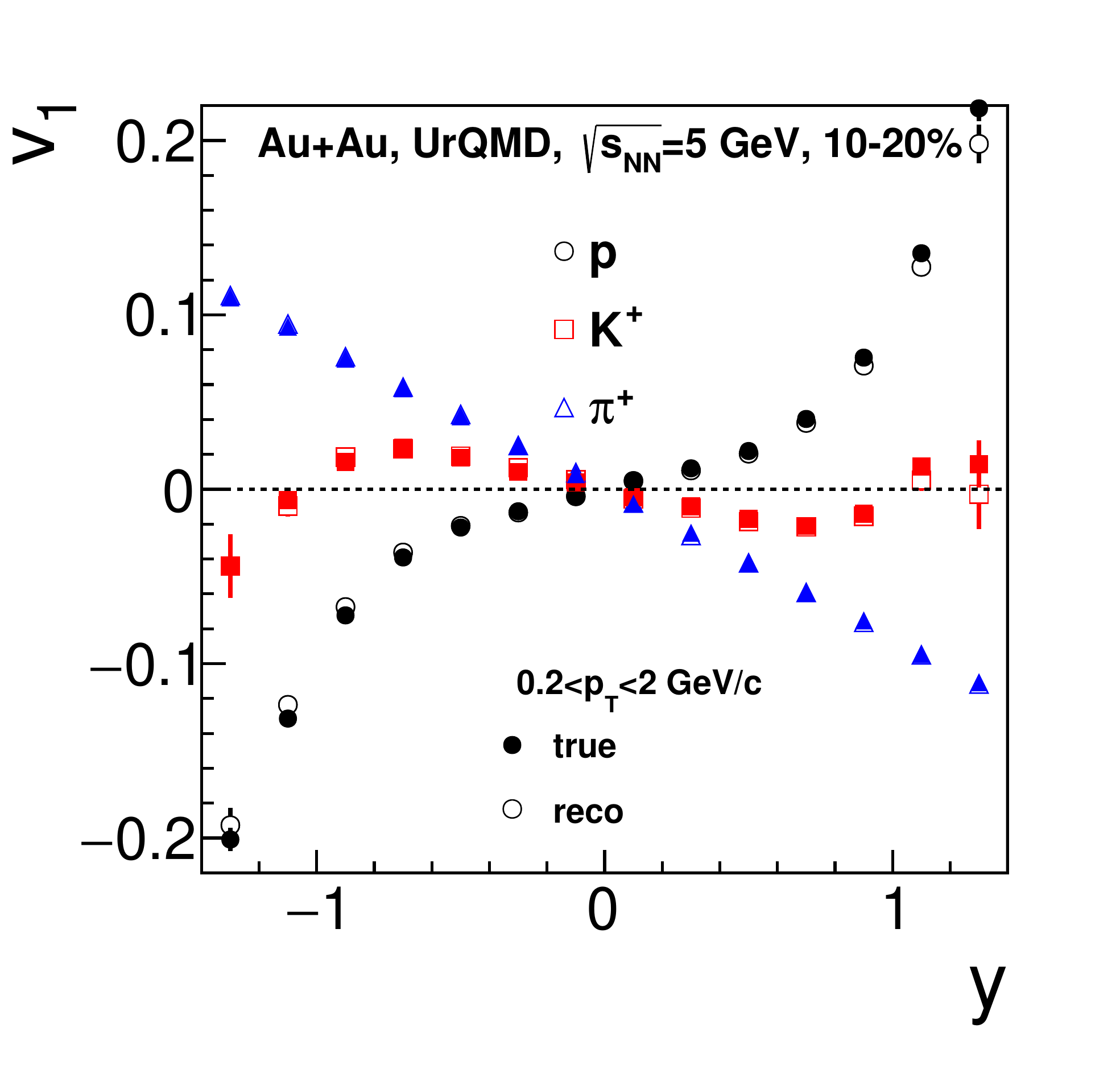}
	\includegraphics[width=0.49\textwidth,clip]{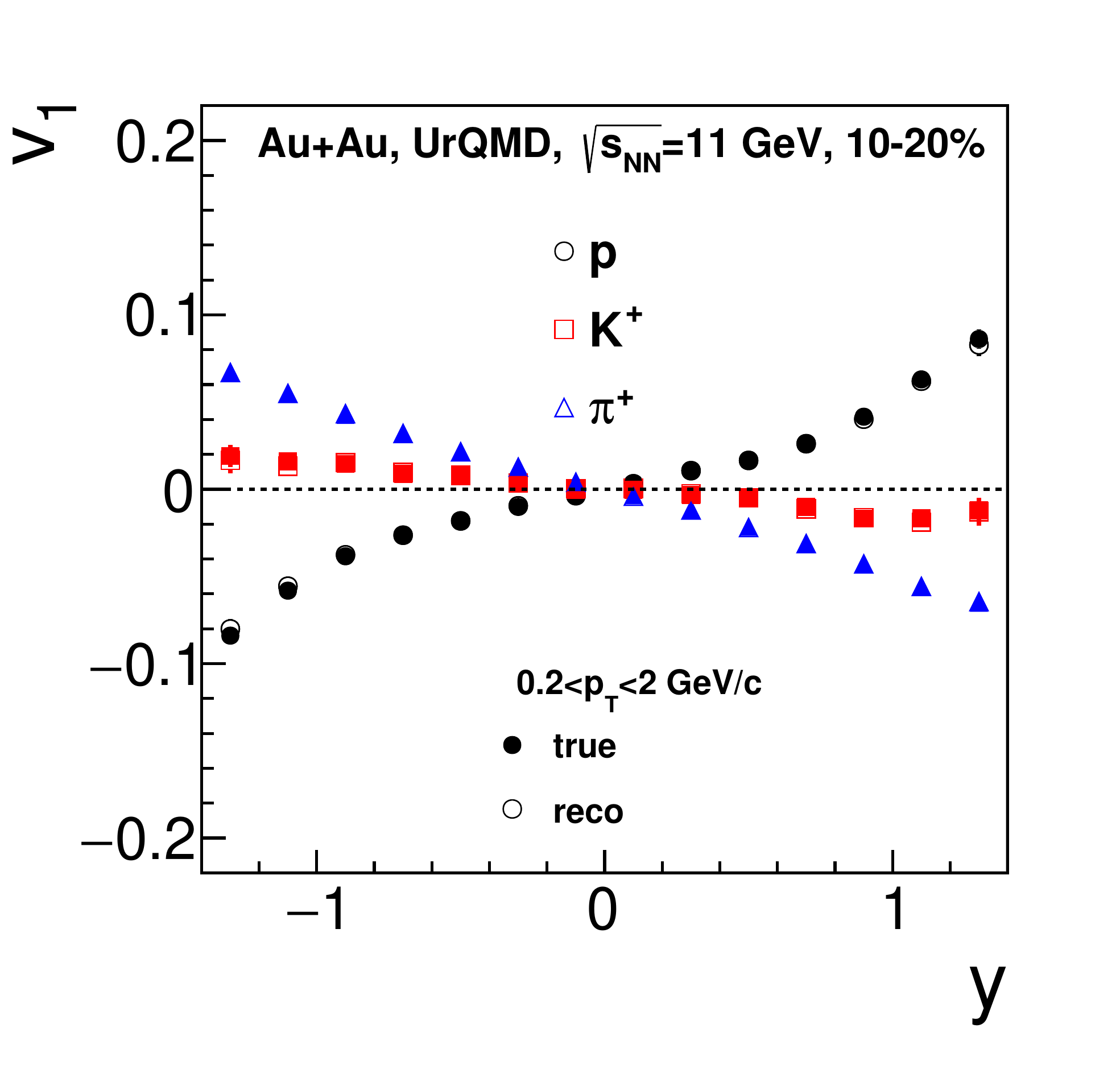}
	\caption{Directed flow $v_1$ as a function of rapidity $y$ for $\sqrt{s_{NN}} = 5$ (left) and $11$ GeV (right). The results from the \GEANT\ simulation are marked as true, and the ones from the reconstruction procedure are marked as reco.}
	\label{fig-v1}       
\end{figure*}

\begin{figure*}
	\centering
	\includegraphics[width=0.49\textwidth,clip]{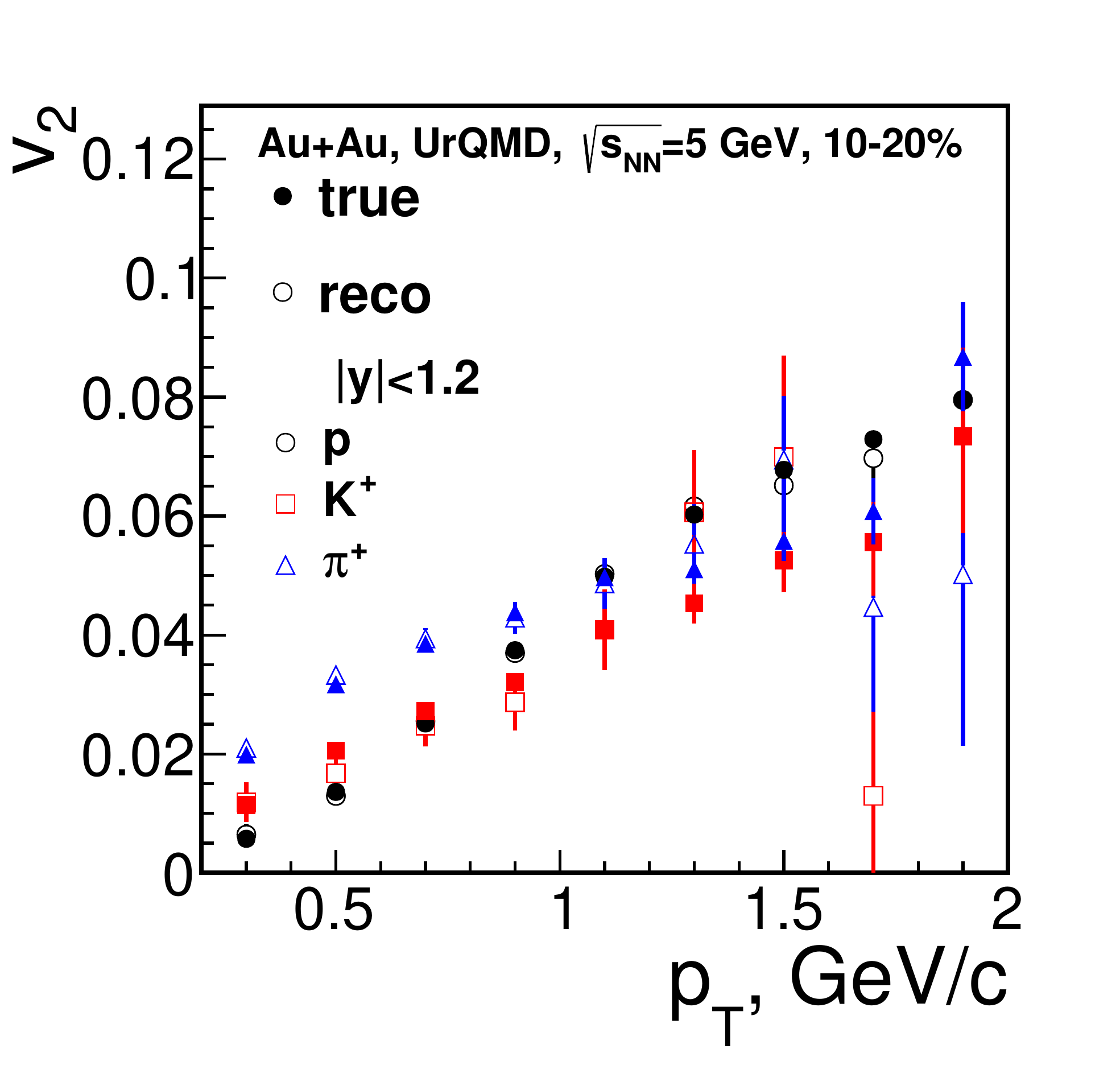}
	\includegraphics[width=0.49\textwidth,clip]{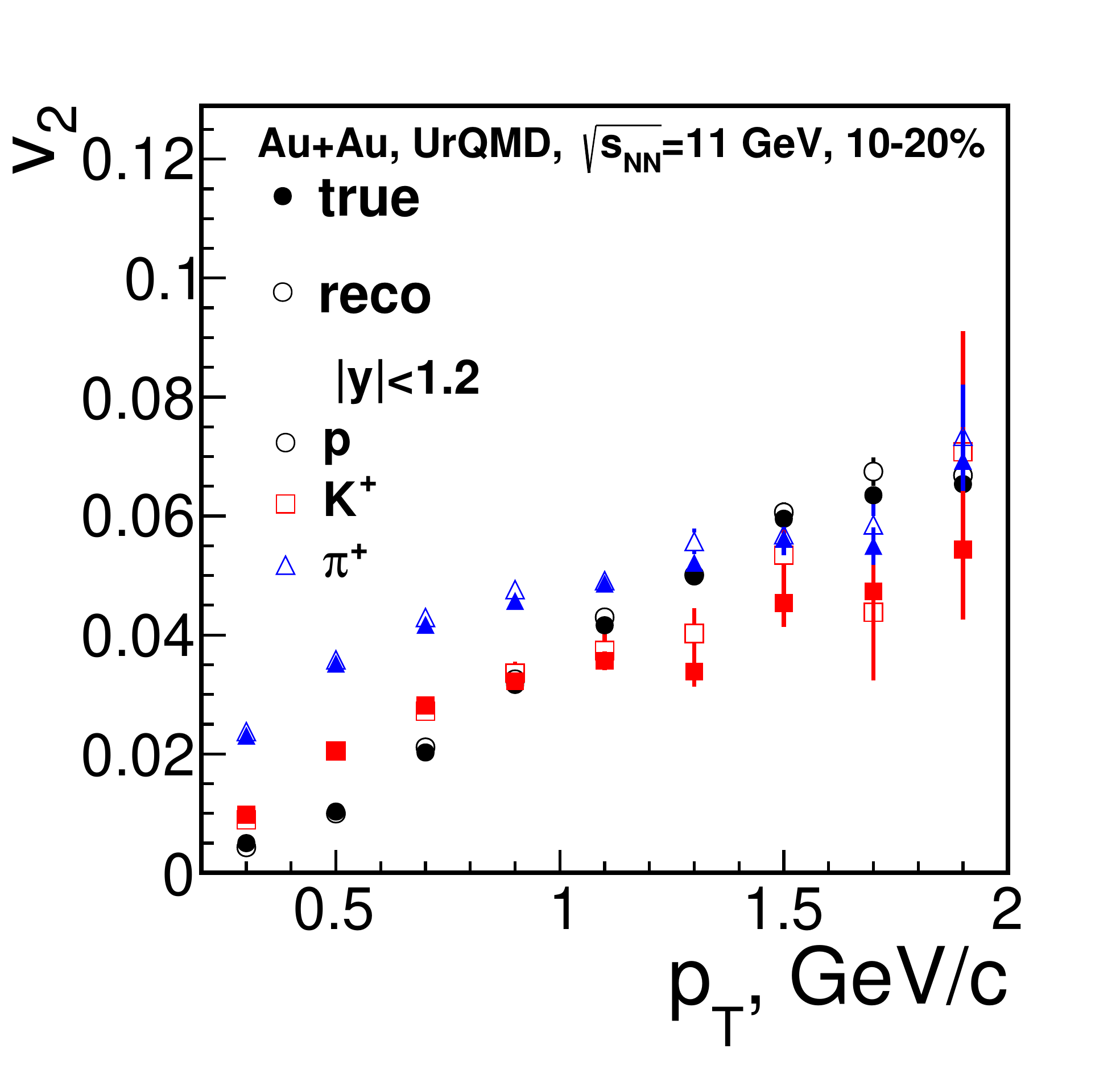}
	\caption{Elliptic flow $v_1$ as a function of transverse momentum $p_T$ for $\sqrt{s_{NN}} = 5$ (left) and $11$ GeV (right). The results from the \GEANT\ simulation are marked as true, and the ones from the reconstruction procedure are marked as reco.}
	\label{fig-v2}       
\end{figure*}

In Figure \ref{fig-v1} one can see the directed $v_1$ flow as a function of rapidity $y$. The plots in Figure \ref{fig-v2} depict the $p_T$-dependence of the elliptic $v_2$ flow.
The results of the \GEANT\ simulations (marked as ''true'') agree with the results from the reconstruction based on the detector response (marked as ''reco''). The azimuthal flow is shown for the \UrQMD\ event generator. Hadrons were identified via a combined \PID\ method \cite{Mudrokh:2017gbk}.

\FloatBarrier
\section{Summary}
\label{sec-summary}

An anisotropic flow performance analysis with full reconstruction chain (which includes realistic particle identification) was shown for the \MPD. The resolution correction factor increases using \GEANT 4 for the highest energy point available at \NICA. The resolution correction factor as well as the directed and elliptic anisotropic flow simulated in \GEANT\ are consistent with the reconstructed values which show a good performance of the \MPD\ detector.

\FloatBarrier
\section{Acknowledgments}
\label{sec-acknowledgments}
This work was partially supported by the Ministry of Science and Education of the Russian Federation, grant N 3.3380.2017/4.6, and by the National Research Nuclear University MEPhI in the framework of the Russian Academic Excellence Project (contract No. 02.a03.21.0005, 27.08.2013).

\FloatBarrier
%

\begin{thebibliography}{}
%
%
%
\bibitem{Czopowicz:2012ey} 
T.~Czopowicz [NA61/SHINE Collaboration],
Prog.\ Theor.\ Phys.\ Suppl.\  {\bf 193}, 29 (2012)
[arXiv:1201.5829 [hep-ex]].
%
\bibitem{Poskanzer:1998yz} 
A.~M.~Poskanzer and S.~A.~Voloshin,
Phys.\ Rev.\ C {\bf 58}, 1671 (1998)
[nucl-ex/9805001].
%
\bibitem{Parfenov:2017rnj} 
P.~Parfenov, I.~Selyuzhenkov, A.~Taranenko and A.~Trutse,
KnE Energ.\ Phys.\  {\bf 3}, 352 (2018)
doi:10.18502/ken.v3i1.1766
[arXiv:1712.09523 [hep-ex]].
%
\bibitem{Svintsov:2017rac} 
I.~A.~Svintsov, P.~E.~Parfenov, I.~V.~Selyuzhenkov and A.~V.~Taranenko,
J.\ Phys.\ Conf.\ Ser.\  {\bf 798}, no. 1, 012067 (2017).
%
\bibitem{Bass:1998ca} 
S.~A.~Bass {\it et al.},
Prog.\ Part.\ Nucl.\ Phys.\  {\bf 41}, 255 (1998)
[Prog.\ Part.\ Nucl.\ Phys.\  {\bf 41}, 225 (1998)]
doi:10.1016/S0146-6410(98)00058-1
[nucl-th/9803035].
%
\bibitem{Bleicher:1999xi} 
M.~Bleicher {\it et al.},
J.\ Phys.\ G {\bf 25}, 1859 (1999)
doi:10.1088/0954-3899/25/9/308
[hep-ph/9909407].
%
\bibitem{Gudima:2001aa} 
K.~K.~Gudima, S.~G.~Mashnik and A.~J.~Sierk,
Los\ Alamos\ National\ Report\  {\bf LA-UR-01-6804} (2001)
doi:10.1016/j.asr.2003.08.057
[nucl-th/0210065].
%
\bibitem{Mashnik:2002uj} 
S.~G.~Mashnik, K.~K.~Gudima, I.~V.~Moskalenko, R.~E.~Prael and A.~J.~Sierk,
Adv.\ Space Res.\  {\bf 34}, 1288 (2004)
doi:10.1016/j.asr.2003.08.057
[nucl-th/0210065].
%
\bibitem{Mudrokh:2017gbk} 
A.~Mudrokh and A.~Zinchenko,
EPJ Web Conf.\  {\bf 138}, 11006 (2017).
doi:10.1051/epjconf/201713811006
%
\end{thebibliography}
%
%

\end{document}